# Low-Complexity Direct Geolocation of Terrestrial GNSS Jammers from Low Earth Orbit

Giacomo Pojani
European Commission
Joint Research Centre
Ispra, Italy
giacomo.pojani@ec.europa.eu

Javier Tegedor
European Commission
Joint Research Centre
Ispra, Italy
javier.tegedor@ec.europa.eu

Joaquim Fortuny-Guasch
European Commission
Joint Research Centre
Ispra, Italy
joaquim.fortuny-guasch@ec.europa.eu

***Abstract*—This paper introduces a low-complexity technique named quasi-direct geolocation (QDG) to perform passive radio-frequency (RF) geolocation of emitters directly in the position domain, akin to direct geolocation (DG) and direct position determination (DPD). The proposed technique drastically reduces the complexity of DG/DPD and is experimentally demonstrated in geolocating a terrestrial jammer at Jammertest 2025 from a repurposed satellite in low Earth orbit (LEO): OPS-SAT PRETTY. The goal of QDG is to enable satellites to contribute to a multi-constellation system for RF interference (RFI) monitoring as opportunistic spectrum sensors in global navigation satellite systems (GNSS) bands, even if these are constrained by low size, weight, and power (SWaP). They can serve as data collectors and/or edge computers. In the former case, QDG can be used to compress large volumes of I/Q samples into minimal signal information, which can be relayed to ground for post-processing via low-capacity downlinks. In the latter case, QDG can be used to compute the geolocation of RFI sources in orbit on low-power on-board computers (OBC). The drawback of these capabilities is lower sensitivity and accuracy than DG/DPD plus limitations on the types of signal sources that can be geolocated, which, nonetheless, include the most common GNSS jammers.**

## I. Introduction

The proliferation of radio-frequency interference (RFI) within and near the frequency bands of global navigation satellite systems (GNSS) has motivated the development and exploitation of passive remote sensing platforms for monitoring the GNSS spectra from space. GNSS spectrum sensors offer unique advantages when mounted on satellites in low Earth orbit (LEO). First, their stand-off distance allows for the reception of terrestrial RFI sources without saturation and with good sensitivity, while preserving the reception of GNSS signals from above. Secondly, their field view covers large swaths of the Earth's surface with short revisit times. Lastly, the propagation effects of the received signals are predictable and traceable to the line-of-sight paths.

Most satellites in LEO are equipped with GNSS receivers for various purposes: positioning navigation and timing (PNT), precise orbit determination (POD), GNSS radio occultation (GNSS-RO), GNSS reflectometry (GNSS-R), etc. When these receivers have the capability to record I/Q samples, they can be repurposed as GNSS spectrum sensors of opportunity for RFI monitoring. In fact, processing these samples can accurately characterize and geolocate RFI sources with algorithms computationally intensive. Therefore, the recorded I/Q samples are relayed to ground and then post-processed with powerful computers. However, this opportunity is bounded by the limited platform resources in space: the throughputs of I/Q samples stress the limits of the on-board storage and the downlink capacity. To cope with these limitations, there are mainly two alternative options: either compressing the data before relay or processing the data in orbit with the on-board computer (OBC), which is normally limited in terms of computational power.

From a broader perspective, any space-based GNSS RFI monitoring system would benefit from integrating as many satellites as possible. This is instrumental to fulfil requirements in terms of coverage, availability, and accuracy of the RFI geolocation service. However, in the current LEO landscape, most space assets are single satellites and fly on platforms that are characterized by low size, weight, and power (SWaP). Cooperative fleets, including tandems, clusters, or the synchronized conjunction of multiple satellite passes, remain relatively rare. Therefore, the effective exploitation of SWaP-constrained satellites flying solo is crucial to the performance of space-based RFI monitoring systems over wide geographical regions. In this landscape, relaxing the platform requirements for RFI geolocation can increase the number of satellites that can participate in a multi-constellation monitoring system.

The work presented in this paper introduces a technique for the low-complexity direct geolocation of GNSS RFI sources, which is named quasi-direct geolocation (QDG). This technique comprises an ensemble of algorithms, which compress and process data to enable near real-time RFI geolocation even on platforms that do not feature high-capacity downlinks or powerful OBCs. The performance of QDG is experimentally demonstrated in a single-pass and single-satellite scenario. Advancing the work in [1], the present study further delves into the I/Q samples collected over Jammertest 2025 by a 3U CubeSat in LEO: OPS-SAT PRETTY [2, 3]. The CubeSat was designed for GNSS-R and later repurposed for other on-demand demonstrations in flight.

## II. State of the Art of Radio-Frequency Geolocation

Traditional techniques of passive geolocation of RF emitters generally comprise two estimation steps [4-10]. In the first step, they measure physical observables of the received signals, such as times of arrival (TOA), frequencies of arrival (FOA), time differences of arrival (TDOA), frequency differences of arrival (FDOA), etc. These time-frequency measurements are estimated from the I/Q samples and aggregated into time series. In the second step, the time series of measurements are fed to nonlinear estimators to solve for the coordinates of a given number of emitters, together with other nuisance parameters. Such two-step techniques suffer from two issues. First, they are inherently sub-optimal because they split the estimation problem into two sub-problems or steps, where the information within the I/Q samples is collapsed into intermediate and independent measurements. Consequently, the accuracy and sensitivity of the position estimates tend to degrade. Secondly, the performance of these techniques is heavily dependent on correctly aggregating a multitude of measurements for each emitter into distinct time series over time and receivers. This aggregation is challenging in the presence of many simultaneous emitters with a variety of modulations and power levels. In fact, ambiguities and artefacts arise in the time-frequency domain from the overlapping of the received signals as well as the correlations both inter-signal and intra-signal. This challenge occurs especially with large numbers of uncooperative emitters, which, remarkably, is the case of RFI sources visible from LEO in the GNSS bands.

To overcome the shortcomings of traditional techniques, single-step techniques were proposed in [11, 12] to geolocate emitters from multiple receivers. The recent publication in [13] demonstrated the application of these techniques to geolocate GNSS RFI sources from two LEO satellites. These are collectively referred to as direct geolocation (DG) or direct position determination (DPD) techniques, because they perform exhaustive search by cross-correlating I/Q samples over a grid of geographic positions. In the position domain, in fact, individual emitters are separated by their physical distance: estimates of their positions can be found around the coordinates of the peaks of the cross-correlation magnitude. This requires no measurement aggregation or prior knowledge of the number of visible emitters.

More specifically, DG/DPD works in one step by cross-correlating I/Q samples over TDOA/FDOA time series, which are parametrized for each single grid point. This skips the intermediate measurements and the relevant issues. Moreover, one-step techniques can outperform two-step techniques in terms of accuracy and sensitivity, which is useful with low signal-to-noise ratio (SNR) and/or short integration times. However, this improvement comes with two disadvantages. The first is a simplifying assumption: the TDOA/FDOA time series are parametrized for an emitter that is stationary within the integration time. The second is a high complexity due to the brute-force nature of exhaustive searches: a search space covering a wide geographic area implies cross-correlating I/Q samples over a massive grid of points. In addition, the longer the integration time is or the wider the emission bandwidth is, the finer the grid. Consequently, DG/DPD imposes a heavy computational burden, which has restricted it to post-processing with graphics processing units (GPU) [14].

In the context of GNSS RFI monitoring from space, powerful GPUs are generally not integrated into SWaP-constrained satellites. They are available on power-hungry computers on ground. Therefore, the DG/DPD of RFI sources relies on relaying large volumes of I/Q samples from multiple satellites in LEO. This downlink may introduce significant latency, which can be reduced only by expanding the geographical distribution and capacity of the network of ground stations.

## III. Jammertest and OPS-SAT PRETTY satellite

Jammertest is regarded as the largest open-air GNSS RFI exercise in the world [3]. It is organized annually in Andøya, Norway, with the coordination of Testnor and several Norwegian public authorities. The added value of this exercise is the exact knowledge of the RFI environment, i.e., position, power, waveform, etc., of all test emitters. Beyond these exercises, most of real-world scenarios are illegal and/or related to military RFI events, without reliable ground truths.

The data examined in this paper comprise I/Q samples collected at Jammertest 2025 as part of a campaign organized by the European Commission’s Joint Research Centre (JRC) in collaboration with the European Space Agency (ESA), the Technical University (TU) of Graz and Testnor. The spacecraft used for the experiment is the Austrian 3U CubeSat OPS-SAT PRETTY (Passive REflecTometry and dosimeTrY), which completed its original mission as GNSS-R demonstrator in 2025 [2]. The satellite integrates a GNSS receiver supporting GPS/Galileo L1/E1 tracking. The RF front end is connected to two Earth-facing patch antennas, which are tuned for GPS/Galileo L5/E5a frequency.

## IV. Quasi-direct RF geolocation

This paper proposes the QDG as a technique to speed up the passive RF geolocation of emitters in the position domain. This technique pre-preprocesses the I/Q samples through a discrete and linear time-frequency transform to adaptively compress the amount of data for direct emitter geolocation. The compression trades accuracy and sensitivity for reduced complexity and restricts the class of signal sources that can be geolocated. This technique is expected to significantly improve over DG/DPD in terms of speed both at the edge, with low-power OBCs, and remotely, through low-capacity links. Both cases are representative of using satellites subject to SWaP constraints as GNSS spectrum sensors. Under the limitations explained in the following, QDG can be used as low-complexity alternative to DG/DPD in [13] to pinpoint terrestrial GNSS jammers from LEO.

Existing studies in [15, 16] propose several techniques to combine the DG/DPD and time-frequency transforms. However, these differ from QDG because they use these transforms to isolate non-stationary waveforms, which are then estimated jointly with the source position for specific classes of signals. Their objective is to enhance the RF geolocation sensitivity of uncooperative emitters in low SNR, rather than lowering complexity. Likewise, among two-step techniques, time-frequency transforms are used for the joint characterization and geolocation of multiple jammers in [17].

### *A. Signal Model*

Let us consider the standard analytic signal model, which adopts the narrowband approximation valid in GNSS bands. For simplicity, let us assume that the signal is transmitted by a stationary emitter. This model over time $\mathrm{t}$ is the complex envelope $\mathrm{s(t)}$ upconverted to the transmit frequency $\mathrm{f_{TX}}$ and baseband bandwidth $\mathrm{B_{TX}}$:

$$\mathrm{x(t)} = \mathrm{s(t)}\, e^{j\,2\,\pi\, f_{TX}\, t},\ \mathrm{f_{TX}} \ll \mathrm{B_{TX}} \tag{1}$$

Let us consider the reception and downconversion of this signal by a moving receiver. If we neglect noise, propagation effects, and the impact of the RF front end, the ideal complex envelope at the $i$-th receiver is:

$$\mathrm{y_i(t)} = \mathrm{s}\left(\left(\mathrm{t} - \tau_i(\mathrm{t})\right)\left(1 - \beta_i(\mathrm{t})\right)\right) e^{j2\pi\left(\Delta f_{LO,i} t + \int_0^t f_{d,i}(\tau) d\tau\right)} \tag{2}$$

The TOA of this signal at the time t is:

$$\tau_i(t) = \frac{\rho_i(t)}{c} + \delta t_{RX,i}(t) - \delta t_{TX}(t) \quad (3)$$

where the time of flight given by the geometric range $\rho_i(t)$ divided by the speed of light c is offset by the receiver clock bias $\delta t_{RX,i}(t)$ and the transmit time $\delta t_{TX}(t)$. The range between the receiver position vector $\mathbf{p}_{RX,i}(t)$ and the static emitter position vector $\mathbf{p}_{TX}$ is:

$$\rho_i(t) = \sqrt{\mathbf{r}_i^T(t)\mathbf{r}_i(t)} \quad (4)$$

with the range vector $\mathbf{r}_i(t) = \mathbf{p}_{RX,i}(t) - \mathbf{p}_{TX}$. The complex envelope $y_i(t)$ is affected by the time dilation/compression factor related to the TOA as $\beta_i(t) = d\tau_i(t)/dt$. If the frequency $f_{RX}$ driven by the receiver local oscillator (LO) does not coincide with $f_{TX}$, the received signal FOA is off by a constant frequency bias $\Delta f_{LO,i} = f_{TX} - f_{RX,i}$. The time-varying component of the FOA of this signal at time t is:

$$f_{d,i}(t) = -f_{TX}\,\beta_i(t) = \quad (5)$$
$$-\frac{f_{TX}}{c}\hat{\mathbf{r}}_i^{\,T}(t)\,\mathbf{v}_i(t) - f_{TX}\left(\delta \dot{t}_{RX,i}(t) - \delta \dot{t}_{TX}(t)\left(1 - \delta \dot{t}_{RX,i}(t)\right)\right)$$

where the physical Doppler shift is proportional to the receiver velocity vector $\mathbf{v}_i(t)$ projected along the range unit vector $\hat{\mathbf{r}}_i(t) = \mathbf{r}_i(t)/\rho_i(t)$, while $\delta \dot{t}_{RX,i}(t)$ and $\delta \dot{t}_{TX}(t)$ are the clock drifts of the receiver and the emitter, respectively.

The received signal model in (2) can be applied to multiple synchronized receivers, which are accurately aligned to the same time reference and tuned to the exact same frequency $f_{RX}$, which implies $\Delta f_{LO,i} = \Delta f_{LO,j}$. Differencing simultaneous TOAs and FOAs at two of such receivers cancels out the transmit time $\delta t_{TX}(t)$ and the emitter clock drift $\delta \dot{t}_{RX,i}(t)$. The resultant TDOAs and FDOAs between the i-th and j-th receivers are:

$$\Delta\tau_{ij}(t) = \frac{\rho_i(t)}{c} - \frac{\rho_j(t)}{c} + \delta t_{RX,i}(t) - \delta t_{RX,j}(t) \quad (6)$$

$$\Delta f_{d,ij}(t) = \quad (7)$$
$$-\frac{f_{TX}}{c}\left(\hat{\mathbf{r}}_i^T(t)\mathbf{v}_i(t) - \hat{\mathbf{r}}_j^T(t)\mathbf{v}_j(t)\right) - f_{TX}\left(\delta \dot{t}_{RX,i}(t) - \delta \dot{t}_{RX,j}(t)\right)$$

Phase differences of arrival (PDOA) between the two receivers may be mathematically expressed as the integral of the FDOAs over time plus the initial phase difference $\Delta\phi_0 = \Delta\phi_{ij}(0) \in [-\pi,\ \pi[$ and as scaled TDOAs plus the integer $N_\phi$ phase ambiguity and a constant phase offset $\Delta\phi_{RX,ij} \in [-\pi,\ \pi[$ due to the different LOs:

$$\Delta\phi_{ij}(t) = 2\pi\int_0^t \Delta f_{d,ij}(\tau)d\tau + \Delta\phi_0 =$$
$$-2\pi\, f_{TX}\Delta\tau_{ij}(t) + 2\pi\, N_\phi + \Delta\phi_{RX,ij} \quad (8)$$

## *B. Signal Pre-Processing*

The first algorithm of QDG pre-processes the I/Q samples through a discrete and linear time-frequency transform. Well known transforms of this family are the short-time Fourier transform (STFT), S-transform and Wavelet transform, which are reviewed in [18]. The STFT projects the signal samples onto the time-frequency domain with a fixed resolution, which is set by the analysis window. This transform can be processed efficiently either in batches through the fast Fourier transform (FFT) or sample by sample through a bank of digital filters.

Let us consider the ideal digitization of the noiseless received signal model in (2) is $y_i[k] = y_i(kT_s)$ with a sampling period $T_s = 1/f_s$. The discrete STFT of the samples $y_i[k]$ at the m-th time bin and n-th frequency bin is:

$$Y_i[m,n] = \sum_{k=0}^{N-1} y_i[k]\, w^*[k - m\,L]e^{-j2\pi\frac{n}{N}k} \quad (9)$$

where $w^*[k]$ is the complex conjugate of the analysis window $w[k]$ sliding over the samples, N is the size of this window coinciding with the number of frequency bins, and L is the hop size or the number of samples by which the window slides. The time resolution $\Delta t = L\,T_s$ and the frequency resolution $\Delta f = f_s/N$ are inversely proportional. If we upper bound the hop size to half the window size, so that all samples in $y[k]$ are used in the processing of (9), the number of time-frequency bins of the entire STFT is always greater than or equal to the number of signal samples:

$$N\,M = N\left(\left\lfloor\frac{K-N}{L}\right\rfloor + 1\right) \geq K,\ \ L \leq \left\lfloor\frac{N}{2}\right\rfloor \quad (10)$$

where K is the number of samples of $y[k]$, $k = 0, \dots, K-1$, and the number M of time bins or hops or windows is derived without zero-padding to handle boundary windows.

## *C. Data Compression*

The STFT projects the I/Q samples into a domain where the most important signal information is concentrated into a few time-frequency bins. Therefore, even though (10) shows that the amount of data grows after pre-processing, this transform can be coupled with efficient algorithms of lossy data compression, which compress the STFT window by window adaptively to the received signals. One simple reference algorithm is to selectively filter out the STFT bins with energy below the noise threshold, assuming that these carry negligible power according to a constant false alarm rate, which tunes the compression ratio.

Compression in the time-frequency domain can be combined with other classical methods of data compression of I/Q samples, such as reducing the bit depth [1] or downsampling after digital downconversion.

Let us assume that the I/Q samples received contain both $y_i[k]$ and additive Gaussian noise $n_i[k]$ with variance $\sigma_n^2$: $z_i[k] = y_i[k] + n_i[k]$. It is proved in [19] that any time-frequency bin in the STFT of $n_i[k]$ can be modeled as a Chi-squared random variable. It follows that we can define a binary decision test to determine the bins that contain meaningful power. Statistically, these bins have energy above the noise threshold:

$$\alpha = -2\hat{\sigma}_n^2 E_w \log P_{FA} \quad (11)$$

where $\hat{\sigma}_n^2$ is an estimate of the noise power, $P_{FA}$ is an arbitrary false-alarm rate, and the energy of the analysis window $w[k]$ is $E_w = \sum_{k=0}^{N-1}|w[k]|^2$.

By squaring the magnitude of the STFT, we can measure the SNR per bin as:

$$SNR_i[m,n] = \frac{|Z_i[m,n]|^2 - 2\hat{\sigma}_n^2 E_w}{2\hat{\sigma}_n^2 E_w} \quad (12)$$

where $Z_i[m,n]$ is the STFT of $z_i[k]$, the estimate of the noise power is $\hat{\sigma}_n^2 \approx \sigma_n^2$.

### *D. General Definition with Two Receivers*

Given two K-size arrays of samples of (2) from two synchronized receivers, their cross-correlation at an arbitrary point with position vector $\mathbf{p}$ over the coherent integration time $T = K\,T_s$ is expressed as:

$$S(\mathbf{p}) = \sum_{k=0}^{K-1} y_i[k] y_j^*\left[k + \Delta\tau_{ij}(\mathbf{p})[k]\right] e^{-j\Delta\phi_{ij}(\mathbf{p})[k]} \quad (13)$$

where $\Delta\tau_{ij}(\mathbf{p})[k]$ and $\Delta\phi_{ij}(\mathbf{p})[k]$ are time series of TDOAs in (10) and PDOAs in (12), respectively, if $\mathbf{p} = \mathbf{p}_{TX}$. For DG/DPD, equation (13) is evaluated at every grid point that is a candidate emitter position [13]. Since this evaluation over large chunks of the grid is computationally demanding, it can be accelerated by exploiting the massive parallel computing power of GPUs [14]. To perform QDG, the computation in (13) is approximated by cross-correlating the pre-processed STFTs of the two arrays of samples $y_i[k]$ and $y_j[k]$:

$$\bar{S}(\mathbf{p}) = \sum_{m=0}^{M-1} \sum_{n\epsilon\bar{N}_m} Y_i[m,n] Y_j^*\left[m + \Delta m_{ij}(\mathbf{p})[m], n + \Delta n_{ij}(\mathbf{p})[m]\right] e^{-j2\pi\Delta\tilde{\psi}_{ij}(\mathbf{p})[m]} e^{-j\Delta\tilde{\phi}_{ij}(\mathbf{p})[m]} \quad (14)$$

where $\bar{N}_m$ denotes the subsets of STFT frequency bins for each m-th time bin, $\Delta m_{ij}(\mathbf{p})[m] = \lfloor \Delta\tau_{ij}(\mathbf{p})[mL]/\Delta t \rceil$ and $\Delta n_{ij}(\mathbf{p})[m] = \lfloor (\Delta f_{d,ij}(\mathbf{p})[mL] + 0.5\, f_s)/\Delta f \rceil$ are the time and frequency bins indices of the TDOAs and FDOAs in (6) and (7), respectively. The PDOAs are proportional to the fractional TDOAs according to $\Delta\tilde{\phi}_{ij}(\mathbf{p})[m] = -2\pi\, f_{TX} \Delta\tilde{\tau}_{ij}(\mathbf{p})[m]$ with $\Delta\tilde{\tau}_{ij}(\mathbf{p})[m] = \Delta\tau_{ij}(\mathbf{p})[mL] - \Delta m_{ij}(\mathbf{p})[m]\Delta t$. The phase is compensated by $\Delta\tilde{\psi}_{ij}(\mathbf{p})[m]$, which is the integral in time of the fractional FDOAs $\Delta\tilde{f}_{d,ij}(\mathbf{p})[m] = \Delta f_{d,ij}(\mathbf{p})[mL] - \Delta n_{ij}(\mathbf{p})[m]\Delta f - 0.5\, f_s$. The use of the STFTs as look-up tables lowers the complexity to evaluate the cross-correlation compared to (13), because it eliminates the need to perform fractional delay interpolation and Doppler shift mixing onto many samples. This speeds up the grid search to the extent that QDG can be performed even sequentially, by processing short chunks of points through the cores of a central processing unit (CPU). The computational demand can be further reduced if data compression algorithms are used to selectively discard bins of the STFTs. This selection scales the subset $\bar{N}_m$ of unfiltered frequency bins over time to capture only the received signals of the sources of interest. The reference algorithm filters out the bins with energy at or below the noise threshold (11) using the false-alarm rate $P_{FA}$ to tune the compression factor.

Resorting to QDG introduces several trade-offs. The use of STFTs implicitly assumes that the signals are quasi-stationary within the duration of the analysis window (i.e., $N\,T_s$). Therefore, in every bin, any significant deviation of the actual signal FDOA or TDOA from the nominal times or frequencies of the bin translates into decorrelation, spectral leakage, and incoherence, which, overall, degrade the SNR and adversely impact the geolocation accuracy.

If a window suitable to the received signals is set for the STFT, the QDG is supposed to effectively geolocate emitters of narrowband modulations, either stationary, such as continuous waves (CW), or frequency swept, such as periodic chirps and frequency-hopped CWs. These are most common waveforms broadcasted by jammers inside GNSS bands. On the contrary, the effectiveness of QDG is expected to be poor in geolocating emitters of wideband noise as well as direct-sequence spread spectrum modulations. These latter are akin to signals transmitted by matched-spectrum GNSS jammers and GNSS spoofers. Besides these limitations, there might be other restrictions on the signals due to data compression. For example, filtering the STFTs through (11) entails that all signals buried in noise are not visible.

### *E. Specific Variant with One Multi-Antenna Receiver*

Direct emitter geolocation based on TOA/FOAs in a single-receiver scenario relies on accurately knowing the waveform of the signal received [1, 5, 11]. However, since the transmit time is generally unknown with sufficient accuracy and the transmitter is uncooperative, in this scenario the DG/DPD and QDG are performed by matching Doppler shifts inside the FOAs at individual antennas and/or through carrier-phase interferometry based on the PDOAs among the elements of an array of antennas. Both methods work effectively with narrowband modulations. In the following, we combine them.

For the sake of simplicity, consider the signal model of a CW: $x(t) = e^{j\,2\,\pi\, f_{TX}\, t}$. This signal arrives to the receiver via an array of two antenna elements indexed by 0 and 1. In such scenario, equation (14) may be rewritten as:

$$\bar{S}(\mathbf{p}) = \sum_{m=0}^{M-1} \sum_{n\epsilon\bar{N}_m} Y_{01}\left[m, n(\mathbf{p})[m]\right] e^{-j\Delta\tilde{\phi}_{01}(\mathbf{p})[m]} \quad (15)$$

where $Y_{01}[m,n] = Y_0[m,n] Y_1^*[m,n]$ is the cross-product of the two STFTs, and $f_d(\mathbf{p})[m] = f_{d,0}(\mathbf{p})[m] \approx f_{d,1}(\mathbf{p})[m]$ so that $n(\mathbf{p})[m] = \lfloor (\Delta f_{LO} + f_d(\mathbf{p})[mL] + 0.5\, f_s)/\Delta f \rceil$ are the frequency bin indices of the FOAs in (5) and $\Delta\tilde{\psi}_{ij}(\mathbf{p})[m] = 0$. Assuming two coherent antennas and one common LO, i.e., $\Delta\phi_{RX,01} = N_\phi = 0$ in (8), the history of the PDOAs between the two elements from a far-field signal source can be approximated with a series of angles of arrival (AOA):

$$\Delta\tilde{\phi}_{01}(\mathbf{p})[m] = -2\pi\, d\, f_{TX}(\rho_0[m] - \rho_1[m])/c \approx 2\pi\, d\, f_{TX} \cos(\theta_{01}[m])/c \quad (16)$$

if $d \ll \rho[m] \approx \rho_0[m] \approx \rho_1[m]$ and where the AOA $\theta_{01}[m] \in [-0.5\pi,\ 0.5\pi[$ implicitly depends on $\mathbf{p}$ and it wraps along the baseline with length d and unit vector $\hat{\mathbf{d}}$:

$$\theta_{01}[m] = \cos^{-1}\left(-\mathbf{r}^T[mL]\, \hat{\mathbf{d}}[mL]/\rho[mL]\right) \quad (17)$$

The coherent integration time is $T \sim (N + (M-1)L)\, T_s$ in (15) and is upper bounded by the STFT time-frequency quantization error, which dominates over physical effects and the RF front end, whereas common-mode errors of the orbit and LO stability are cancelled out between the two antennas. Performing single-satellite QDG requires evaluating (14) over FOA/PDOA time series, which may span multiple and disjoint time intervals. Cross-correlations can be accumulated semi-coherently, by summing the magnitudes of (15) across these time intervals. Assuming that the noise power is constant over the whole semi-coherent integration time and that the combined antenna gains are constant over the grid of points, the resultant sum in the position domain is scaled to SNR, in the form of deflection coefficient, according to:

$$SNR(\mathbf{p}) = \frac{\sum_{p=0}^{N_{NC}-1}\left(|\bar{S}_i(\mathbf{p})|^2 - \sum_m^{M_i-1} \sum_{n\epsilon\bar{N}_m} \gamma_m(\mathbf{p})[m]\right)}{\sqrt{\sum_{i=0}^{N_{NC}-1}\left(\sum_m^{M_i-1} \sum_{n\epsilon\bar{N}_m} \gamma_m(\mathbf{p})[n]\right)^2}} \quad (18)$$

$$\gamma_m(\mathbf{p})[n] = 4\hat{\sigma}_n^4 E_w^2 \left(\frac{|Y_0[m,n(\mathbf{p})[m]]|^2}{2\hat{\sigma}_n^2 E_w} + \frac{|Y_1[m,n(\mathbf{p})[m]]|^2}{2\hat{\sigma}_n^2 E_w} - 1\right) \quad (19)$$

## V. Experimental Results

The dataset for the experimental results was collected with OPS-SAT PRETTY over Jammertest 2025. Details about the satellite and the collection strategy may be found in [1, 3]. One stationary test jammer was placed on ground around the village of Bleik near Andøya and stably radiating 50 W of effective isotropic power at both L1 and L5 frequencies. The I and Q samples were acquired at 5 Msps with an effective depth of 12 bits and stored as 16-bit integers. The results presented focus on the most favorable of the four passes of OPS-SAT PRETTY. This pass lasted 12 min starting on September 12th around 11:31 UTC with a maximum elevation of 50° above Bleik (see Fig. 1). The satellite made 27 acquisitions of 1 s each, every 9 s, storing about 1 GB of I/Q samples with both antennas in L5 alone. The samples were accompanied with logs of the on-board GNSS receiver and the attitude determination and control system (ADCS). The downlink of the dataset for the whole campaign, i.e., ~8 GB, occupied the ground station at TU Graz for weeks.

OPS-SAT PRETTY is designed for GNSS-R and repurposed for RFI monitoring in this experiment. The LO is not disciplined by the GNSS receiver. A constant frequency offset of $\Delta f_{LO} \approx 8.11$ kHz at L5 is compensated with respect to the FOAs expected from the test jammer true position. The residual error in frequency due to STFT quantization, LO, orbit, etc. is estimated to be ~8 Hz rms with a rate of ~3 Hz/s rms. The antenna array baseline is supposed to be in line with the along-track direction. Comparing the satellite velocity by the GNSS receiver and the quaternions from the ADCS show an attitude error below 1° rms. Since the antenna array is not calibrated, an unknown phase offset exists between the two patch elements. As this offset is constant over the short integration time, it does not reduce the SNR. The baseline spans 105 mm, which is about 0.41 the wavelength of L5. This means that the PDOAs have a range cramped to ±148° but do not wrap around the baseline. Therefore, the PDOA is more sensitive to errors for AOAs from the sides of the array and more precise for AOAs from boresight. Luckily, the expected AOAs of the test jammer range from 26° to 156°. The resultant PDOA error is estimated to be ~7° rms.

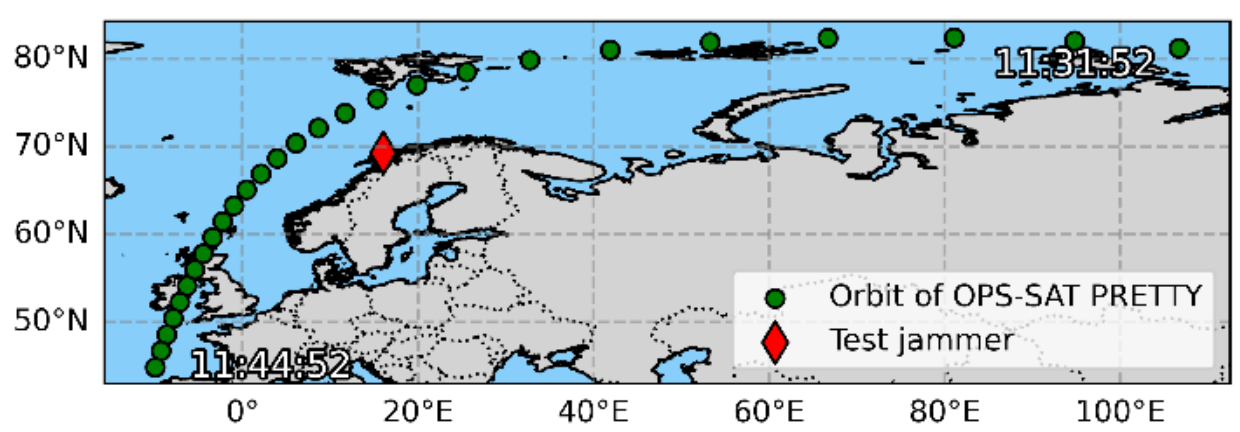

Fig. 1. GNSS positions of the satellite every 30 s during the 12-min pass.

The SNR (17) of one of the two compressed STFT is plotted in Fig. 2, using Hann windows with 50% overlap and resolutions of $\Delta f \approx 10$Hz and $\Delta t \approx 52$ ms. This spectrogram shows the CW at L5 from Bleik as well as other low-SNR signals. The noise power estimate is $2\hat{\sigma}_n^2 E_w = -10.43$ dBFS.

The data memory size changes along the processing chain of QDG as listed in Table I. The last stage is the cross-product between the two compressed STFTs, i.e., $Y_{01}[m,n]$, which is computed before the phase rotation and coherent integration in (15) and is the minimal signal information to relay for post-processing. The highest compression factor of 99.88% is achieved after resampling at 312.5 ksps to extract any CWs with Doppler shift sweeping between ±30 kHz from the L5 carrier. Without resampling, the compression factor is 99.93%. Note that these factors are plausible because the bins with energy above noise are just ~0.08% of the total.

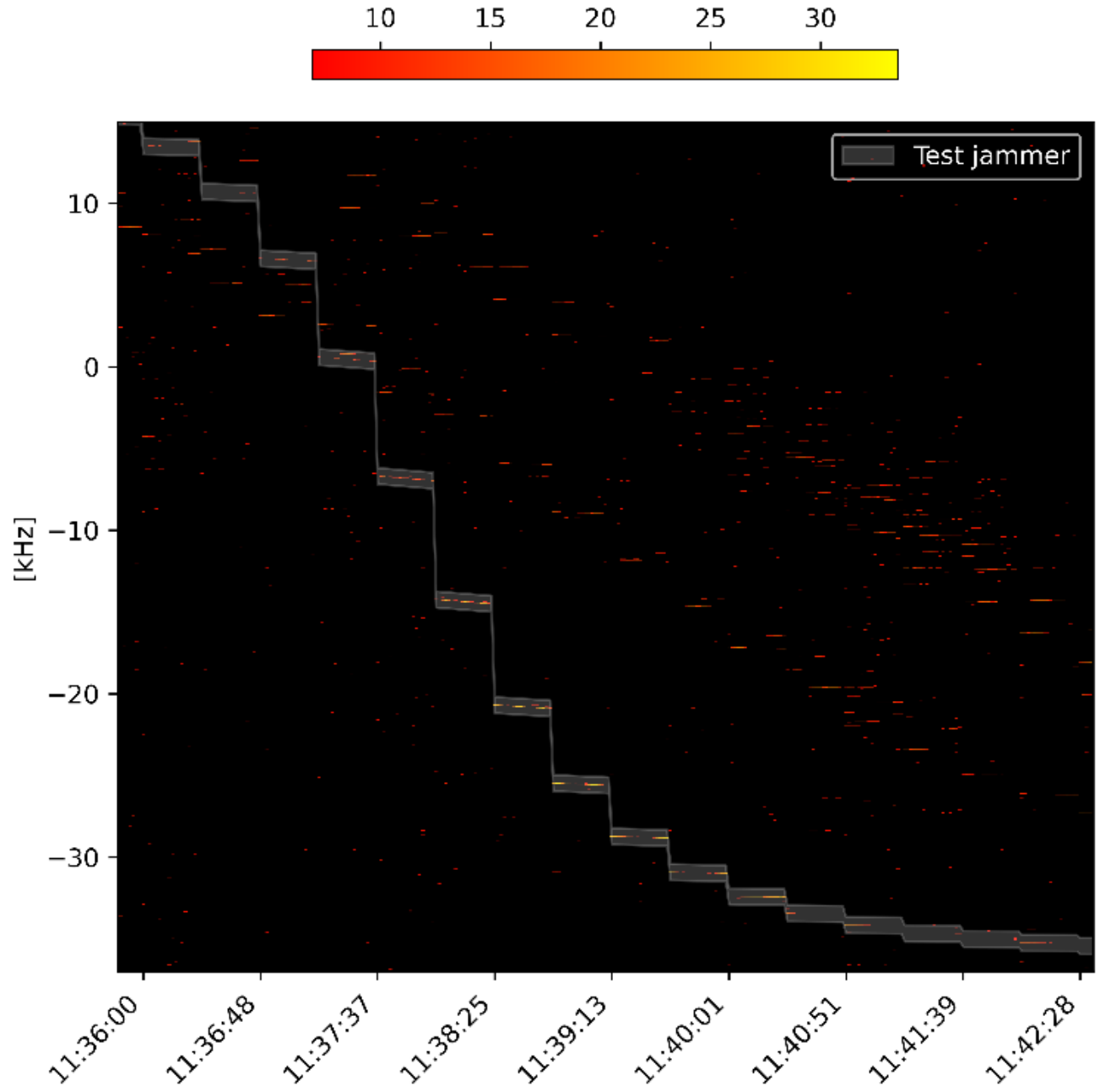

Fig. 2. Close-up view on the SNR spectrogram above the noise threshold in (11), i.e., 8.4 dB with $P_{FA} = 0.001$, for the front patch antenna at L5.

TABLE I. Signal information along QDG processing chain

| Data | Memory size [MB] | |
|---|---|---|
| Collected I/Q samples (2 x int16) | 1080.0 | |
| ***Resampling from 5 Msps to 312.5 ksps*** | ***Yes*** | ***No*** |
| Resampled I/Q samples (2 x int16) | 67.50 | 1080.0 |
| Full STFTs (complex64) | 254.80 | 4076.86 |
| Compressed STFTs[a,b] (complex64 + int32) | 0.29 | 4.63 |
| Product of STFTs[b] (complex64 + int32) | 0.08 | **0.76** |

a. According to noise threshold in (11) with $P_{FA} = 0.001$.
b. Including the indices of the unfiltered time-frequency bins.

Consider the cross-product $Y_{01}[m,n]$ across all the 27 1-s acquisitions made by OPS-SAT PRETTY. After compression, this product occupies 0.76 MB of memory (see Table I). We then use it to compute the SNR in (18) of the semi-coherent combination of 1-s coherent integrations (15) over two grids of WGS84 latitudes and longitudes: one grid in Fig. 3 is wide and coarse, the other one in Fig. 4 finely covers Andøya. Both are centered on the true jammer position [1]. Since the test jammer is on ground, the grid altitudes are set at heights that are interpolated from the global digital elevation model (GDEM) in [20], which is regridded at 0.0625° to occupy 26 MB. The large spacing over the undulated terrain of Andøya causes height errors. Therefore, instead of the hard constraint in [4], the estimation is soft-constrained to the GDEM with a standard deviation of 300 m. Accounting for the estimated receiver position, velocity and time (PVT) covariance, the Cramer-Rao lower bound yields a 95th percentile isotropic circular error probable (CEP95) of ~900 m at the true position. The SNR maximum peak is contained within this bound and, remarkably, lies at the center in both grids. This means the estimate falls within the 0.5-km² area around the ground truth. These grids have sizes of ~125000 points and require computation times on the order of tens of seconds with one single CPU core.

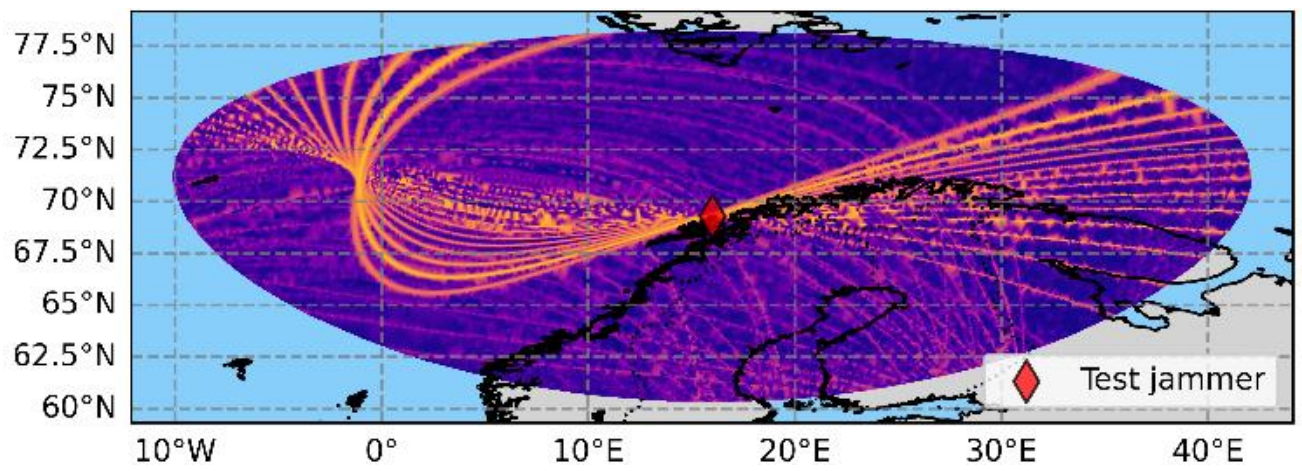

Fig. 3. SNR [dB] over a grid with radius of 1000 km and resolution of 5 km (the mirror lower peak on the West is due to the ambiguity of FOAs/PDOAs).

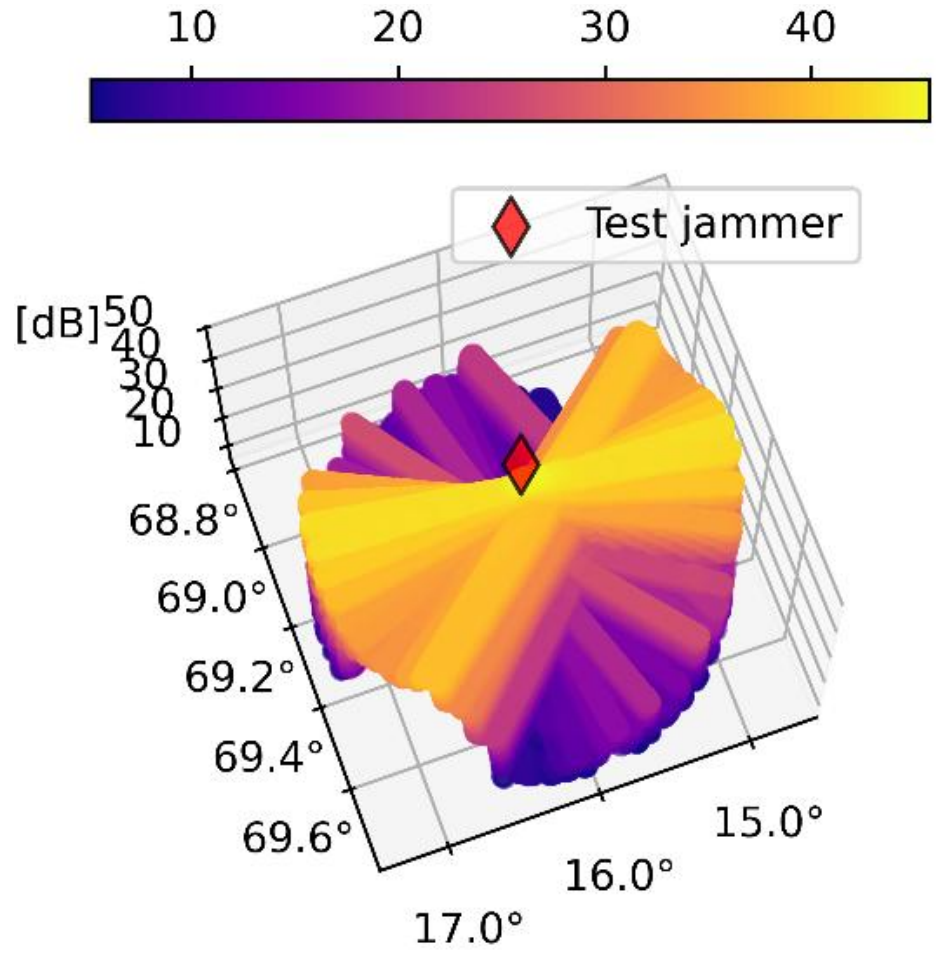

Fig. 4. SNR [dB] over a grid of 50 km with resolution of 250 m (the values range from 5 dB to 45 dB with maximum at the true test jammer position).

## VI. SUMMARY AND CONCLUSIONS

This work proposes and demonstrates the use of QDG for the low-complexity geolocation of GNSS jammers from LEO. The results presented for a single-satellite scenario show that the source of a 50-W CW at L5 can be pinpointed within 0.5 km² by integrating less than 1 MB of data through QDG. These data compress 1 GB of I/Q samples collected in 12 min during one pass of OPS-SAT PRETTY. Their processing with QDG takes seconds with one single CPU core.

The goal of QDG is to allow satellites of opportunity, which have limited computational and downlink capacities, to be repurposed for GNSS RFI monitoring. For instance, QDG can be performed efficiently in orbit with low-power CPUs to compute quick looks of the RF energy on the Earth's surface, to accurately geolocate RFI hotspots in the GNSS bands. Compressed data of quick looks from multiple satellites and/or passes can be relayed with low latency to the ground, where they are combined to deliver increased geolocation accuracy of GNSS jammers and to keep track of RFI activity.

## ACKNOWLEDGMENT

The experiments with OPS-SAT PRETTY during Jammertest 2025 were made possible with the collaboration of ESA, TU Graz, Testnor, and the Norwegian Defense Research Establishment (FFI). Special thanks to David Evans and Tim Oerther from ESA, Maximilian Henkel from TU Graz, and Jonas Stene Lindbjør from Testnor.